\documentclass[sigconf]{acmart}
\newcommand{\HowManyCol}{2}

\AtBeginDocument{%
  }

\copyrightyear{2026}
\acmYear{2026}
\setcopyright{cc}
\setcctype{by}
\acmConference[CHI '26]{Proceedings of the 2026 CHI Conference on Human Factors in Computing Systems}{April 13--17, 2026}{Barcelona, Spain}
\acmBooktitle{Proceedings of the 2026 CHI Conference on Human Factors in Computing Systems (CHI '26), April 13--17, 2026, Barcelona, Spain}
\acmPrice{}
\acmDOI{10.1145/3772318.3791556}
\acmISBN{979-8-4007-2278-3/2026/04}

\newcommand{\papertitle}{Framing Responsible Design of AI for Mental Well-Being:\\AI as Primary Care, Nutritional Supplement, or Yoga Instructor?}
\newcommand{\paperkeywords}{Responsible Artificial Intelligence, Design, Mental Health, Large Language Models} %
\usepackage{enumitem}

\usepackage{colortbl,acmart-taps} 
\usepackage{rotating} 
\usepackage{array}
\usepackage{multirow, makecell}
\usepackage[usestackEOL]{stackengine} 
\usepackage{siunitx}
\usepackage{graphicx}
\usepackage{dblfloatfix}
\usepackage{xspace}
\usepackage{hyperref} 
\usepackage{tabularx} 
\usepackage{bbding} 
\usepackage{MnSymbol}

\definecolor{mygray}{rgb}{0.94,0.94,0.94}

\newcommand{\myparagraph}[1]{\vspace{0.2em}\noindent\textbf{\textit{#1}}\hspace*{.3em}}
\aptLtoXcmd{}{\renewenvironment{quote}
  {\list{}{\rightmargin=0.5cm \leftmargin=0.5cm}%
   \item\relax}
  {\endlist}}

\newcommand{\ifOneColThen}[1]{\ifthenelse{\equal{\HowManyCol}{1}}{#1}{}}

\newcommand{\ifTwoColThen}[1]{\ifthenelse{\equal{\HowManyCol}{2}}{#1}{}}

\newcounter{enumi_saved}

\begin{document}

\title{\papertitle}


\author{Ned Cooper}
\email{ned.cooper@cornell.edu}
\orcid{0000-0003-1834-279X}
\affiliation{
  \department{Information Science}
  \institution{Cornell University}
  \city{Ithaca}
  \state{NY}
  \country{USA}
}

\author{Jose A. Guridi}
\email{jg2222@cornell.edu}
\orcid{0000-0003-0543-699X}
\affiliation{
  \department{Information Science}
  \institution{Cornell University}
  \city{Ithaca}
  \state{NY}
  \country{USA}
}

\author{Angel Hsing-Chi Hwang}
\email{angel.hwang@usc.edu}
\orcid{0000-0002-0951-7845}
\affiliation{%
  \department{}
  \institution{University of Southern California}
  \city{Los Angeles}
  \state{CA}
  \country{USA}}

\author{Beth Kolko}
\email{bkolko@uw.edu}
\orcid{0000-0003-2949-2726}
\affiliation{%
  \department{Human Centered Design \& Engineering}
  \institution{University of Washington}
  \city{Seattle}
  \state{WA}
  \country{USA}
  }

\author{Emma Elizabeth McGinty}
\orcid{0009-0003-5095-9308}
\email{emm4010@med.cornell.edu}
\affiliation{%
  \department{Division of Health Policy and Economics}
  \institution{Weill Cornell Medicine}
  \city{New York}
  \state{NY}
  \country{USA}
  }

\author{Qian Yang}
\email{qianyang@cornell.edu}
\orcid{0000-0002-3548-2535}
\affiliation{%
  \department{Computing and Information Science}
  \institution{Cornell University}
  \city{Ithaca}
  \state{NY}
  \country{USA}
}


\begin{abstract}
Millions of people now use non-clinical Large Language Model (LLM) tools like ChatGPT for mental well-being support.
This paper investigates what it means to design such tools responsibly, and how to operationalize that responsibility in their design and evaluation.
By interviewing experts and analyzing related regulations, we found that designing an LLM tool responsibly involves: (1) Articulating the specific benefits it guarantees and for whom. Does it guarantee specific, proven relief, like an over-the-counter drug, or offer minimal guarantees, like a nutritional supplement? (2) Specifying the LLM tool's ``\textit{active ingredients}'' for improving well-being and whether it guarantees their effective delivery (like a primary care provider) or not (like a yoga instructor).
These specifications outline an LLM tool's pertinent risks, appropriate evaluation metrics, and the respective responsibilities of LLM developers, tool designers, and users.
These analogies—\textit{LLM tools as supplements, drugs, yoga instructors, and primary care providers}—can scaffold further conversations about their responsible design.
\end{abstract}



\maketitle

\section{Introduction}

For better or worse, people are increasingly using non-clinical Large Language Model (LLM) tools, such as ChatGPT and Replika, for mental well-being and self-care~\cite{WSJ_kids_chatbot_as_therapists,Scholich2025Comparison}.
Some specifically target well-being. Others serve broader purposes. These non-clinical tools are intended for users who do not have diagnosed mental illnesses or suicidal thoughts, and can be used without a clinician's prescription or supervision.

Non‑clinical LLM tools promise significant benefits, but also pose real risks.
They can enhance healthy individuals' self-care and comfort the ``\textit{worried well}''~\cite{wagner1984health}, thereby alleviating severe shortages of mental healthcare providers in the U.S.~\cite{apa_burnout_need_for_MH_care,apa_news_need_for_MH_care,apa_survey_need_for_MH_care,white_house_need_for_MH_care,nih_burnout_need_for_MH_care,torous2016digital}.
However, ongoing lawsuits allege these tools have contributed to user suicides~\cite{characterai-suicide-lawsuit-2024,Hill2025A,SMVLC2025Smvlc}. 
User interviews suggest LLM consultations can delay clinical help-seeking and foster oversimplified views of mental illness~\cite{YangGangDIS25_Identity_writing,YangGangCHI24_MAMH, Glossy_easy_futures,goopification2023-Atlantic}.
A randomized controlled trial found that participants who used ChatGPT more showed increased loneliness and reduced socialization~\cite{chatGPT_loneliness_reduce_socialization_MIT2025}.
Finally, scholars warn of potential additional harms that are likely but not yet detected, such as LLM tools worsening health inequalities~\cite{ong2024artificial}.

We are HCI designers and researchers who create non-clinical LLM tools for mental well-being. Acutely aware of these tools' promises and risks, we wanted to translate this awareness into design and evaluation actions. We wanted to design these tools \textit{responsibly}---but how?
LLM tools' benefits and risks for mental well‑being are numerous, still unfolding, and hard to measure objectively or quickly~\cite{AI-self-presentation-CHI23EA,torous2016digital}. Which of these benefits and risks should we prioritize and act on, and how?

This paper investigates what it means to design non-clinical LLM tools for mental well-being responsibly, and how to operationalize that responsibility in their design and evaluation. Little research has directly addressed these questions.

Given the limited prior work in this area, we chose to investigate these questions by posing them directly to experts in related areas.
We interviewed $24$ experts, each with over five years of experience in areas such as responsible AI in digital therapeutics, medical ethics, AI regulation, and AI-related health policies.
As interviewees converged on health policy frameworks to frame the responsible design of LLM tools, we collected and analyzed over $100$ relevant policy documents to contextualize these frameworks, before confirming our findings with the interviewees.


This research process revealed three key interrelated findings:

\begin{enumerate}[leftmargin=*]
    \item What it means to design a LLM mental well-being support tool responsibly depends on the specific benefits it \textit{guarantees} to deliver and for whom.
    Is the tool accountable for providing specific relief to individuals with specific symptoms or diagnoses (like an over-the-counter drug), or does it promise to improve the general well-being of healthy individuals, without guaranteeing any immediate results (like a nutritional supplement)?
    The answer to this question was central to how the experts we interviewed identified the tool’s primary responsibilities and risks, and how they distinguished the responsibilities of the tool's creators from those of its users.

    \item To design an LLM tool responsibly is to articulate its ``\textit{active ingredients}" (\textit{i.e.}, the proven mechanism through which it will deliver the guaranteed health benefits) and ensure their effective delivery.
    Consider an LLM tool that provides Cognitive Behavioral Therapy (CBT). Is the tool accountable for the effective delivery of the treatment (like a primary care provider), or is it more like a yoga instructor, offering no guarantees regarding the health benefits a student will gain from the exercise? Our expert interviewees viewed the former as the more responsible choice.

    \item Interviewees agreed that designing an LLM tool responsibly requires ensuring its risks are commensurate with its guaranteed benefits. However, they sometimes disagreed on how to define ``commensurate'' risks and benefits.
    For instance, some viewed nutritional-supplement-like LLM tools (minimal guaranteed health benefits, minimal safety risks) as responsible; others disagreed. 
    If an LLM therapy bot---analogous to a breakthrough drug---can cure a severe, otherwise untreatable mental illness but carries life-and-death risks for a small subset of users, some considered it responsible; others expressed serious reservations.

\end{enumerate}

These findings are valuable in two ways. First, they reveal actions that designers and researchers can readily take to design LLM mental well-being support more responsibly, namely, (i) articulating specific guaranteed benefits for intended users, (ii) guaranteeing effective delivery of proven active ingredients, and (iii) ensuring commensurate risks and benefits.

Second, the analogies this research offers—\textit{non-clinical LLM tools as nutritional supplements, over-the-counter medications, yoga instructors, and primary care providers}—can serve as a scaffold for further deliberations and debates about their responsible design in future HCI research.
For example, should we evaluate LLM tools based on population-level benefits, even if that means accepting life-or-death risks for a small subset of users, as regulators sometimes do with breakthrough drugs? Do we, HCI communities, endorse yoga-instructor-like and primary-care-like LLM tools as equally responsible designs?

\section{Related Work}

\emph{Content warning: This chapter includes discussions of suicidal ideation.}

\subsection{LLM Mental Well-Being Support in Research}
\label{RW-risk-list}

Recent decades have seen AI incorporated into two distinct types of software: \textit{clinical systems for mental health}, intended for diagnosing and treating mental illnesses, and \textit{non-clinical tools for mental well-being}, intended for the general well-being of those without diagnosed mental illnesses.
U.S. law reinforces this functional distinction: The former are strictly regulated as medical devices by the Food and Drug Administration (FDA). The latter are regulated relatively lightly by the Federal Trade Commission (FTC) as consumer technology.

In practice, the benefits and risks of clinical and non-clinical systems overlap.
Mental well‑being support tools can meaningfully improve users' mental health and well-being. For instance, they can reduce the stigma around mental health, making patients more likely to seek clinical treatments~\cite{reategui2025llm}.
They can provide a listening ear to the ``\textit{worried well}''~\cite{torous2016digital,wagner1984health}, freeing up limited mental‑healthcare resources for those in greater need~\cite{apa_burnout_need_for_MH_care,apa_news_need_for_MH_care,apa_survey_need_for_MH_care,white_house_need_for_MH_care,nih_burnout_need_for_MH_care}.
They can also offer prompt, low-intensity care to vast populations, thereby alleviating the social contagion of mental health conditions~\cite{knapstad2020effectiveness}. 

On the other hand, mental well-being support tools can pose substantial risks and cause real harms to users’ health.
As tools like ChatGPT and Replika are increasingly used (or misused) for mental well-being support, these risks have garnered substantial attention.
Specifically, empirical research has identified the following risks:

\begin{itemize}[leftmargin=*]
    \item \textit{Rare but severe safety risks:~}An ongoing lawsuit alleges that an LLM chatbot, \texttt{character.ai}, encouraged a teenage user to attempt suicide; the user subsequently died~\cite{characterai-suicide-lawsuit-2024}.~Similar lawsuits are underway against OpenAI concerning GPT-4o~\cite{Hill2025A,SMVLC2025Smvlc}.
    \item \textit{Risks of displacing necessary self-care (\textit{e.g.}, socialization, meaningful human relationships).~}A randomized controlled trial found that higher daily ChatGPT usage correlated with increased loneliness and reduced socialization~\cite{chatGPT_loneliness_reduce_socialization_MIT2025}. A recent user interview study echoed this finding~\cite{YangGangDIS25_Identity_writing}.
    \item \textit{Risks of rendering an overly rosy public perception of mental illnesses and mental healthcare}. An interview study suggested that digital mental health apps ``\textit{sell glossy, easy futures}'' to users~\cite{Glossy_easy_futures}. Relatedly, experts have warned that LLM chatbots are ``\textit{goopifying}'' mental healthcare~\cite{goopification2023-Atlantic}. 
    \item \textit{Risks of causing emotional discomfort or distress to users.~}
    An analysis of GPT-4o responses found the model sometimes expressed stigmatizing language toward people with mental health conditions~\cite{Moore_LLM_express_stigma_FAccT2025}; A survey found that LLM chatbot users occasionally perceived bot responses as intrusive and lacking empathy~\cite{chandra2025lived}. 
\end{itemize}

In addition to the risks identified in empirical research, scholars also warn that LLM mental well-being support may cause additional harms that have not yet manifested or been measured.

\begin{itemize}[leftmargin=*]
    \item \textit{Risks of inducing psychosis.~}LLM-powered chatbots can hallucinate and be overly sycophantic toward users~\cite{openaiSycophancyGPT4o, Cheng2025Elephant}. Many online articles link this sycophancy to ``\textit{AI-chatbot-induced psychosis}''~\cite{statnewsReportsPsychosis}.
    However, no research study has yet proven this effect (or clearly defined it~\cite{pbsWhatKnow}) at the time this paper is submitted.
    \item \textit{Risks of replacing or delaying users' clinical care seeking}~\cite{YangGangCHI24_MAMH}. A 2023 survey of 1,000 Americans showed that 80\% perceived ChatGPT as ``\textit{an effective alternative to therapy}''~\cite{tebraPerceptionsHealthcare}.
    \item \textit{Risks of exacerbating inequitable access to care}~\cite{badr2024digital,hatef2024development}, because not all patients have equal access to LLMs, and not all institutions can benefit equally from fine‑tuning models~\cite{obradovich2024opportunities}.
\end{itemize}

How can designers create and evaluate LLM mental well-being support tools responsibly, such that the tools measurably deliver the aforementioned benefits while minimizing these various risks?
To our knowledge, no HCI research has yet directly addressed the question.

This research gap not only affects industry practice (more on this in the next section) but also hinders HCI researchers from designing responsible and measurably beneficial LLM tools. For example, to expand the reach of mental healthcare, HCI researchers have developed valuable LLM-based tools that provide low-intensity care for generally healthy users (\textit{e.g.}, tools that assist reflective and positive self-narrative writing~\cite{ExploreSelf_Song_CHI25,sharma-etal-2023-cognitive}, or offer motivational interview counseling \cite{GPTCoach_JorkeCHI25}).
However, demonstrating that these tools achieve their goal of expanding healthcare reach or that they do not cause potential harms (\textit{e.g.}, an overly rosy public perception of mental wellness) remains challenging. As a result, researchers have often evaluated these tools using measures such as mental health diagnostic questionnaires or user-perceived usefulness, but rarely incorporate measures of societal impact.

\subsection[Attempts to Design AI Mental Well-Being Support Responsibly]{Industries' Attempts to Design AI Mental Well-Being Support Responsibly}
\label{RW-design-action-challenges}


Companies have pursued two main approaches to design non-clinical AI tools for mental well-being more responsibly and demonstrate their accountability. The first approach is to design and evaluate these non-clinical tools to meet clinical standards, \textit{i.e.}, to obtain FDA (Food and Drug Administration) approval.
When a mental well-being support tool demonstrates to the FDA that it meets medical device safety and effectiveness standards across its target populations, it can reassure users and the public that these two risks---arguably the most important on the risk list of \S\ref{RW-risk-list}---are adequately addressed.


Under current U.S. regulations, however, meeting clinical standards has proven extremely challenging, if not impossible, for non-clinical AI tools.
Consider Woebot, a pioneer in therapy chatbots. Woebot used natural language processing (NLP) to interpret user utterances and a simple rule-based system to deliver personalized CBT.
Once operated in over 120 countries, Woebot ultimately shut down while seeking FDA approval, in part because ``\textit{the FDA had not yet figured out how to regulate}'' such a system, and in part due to competition from LLM-based tools that did not pursue FDA approval~\cite{Woebot_shutdown2025,kahane2025policy}.
Woebot’s fate was not unique. Before Woebot, most companies seeking FDA approval for non-AI mental health tools, such as video games for people with ADHD, went bankrupt during or shortly after the approval process~\cite{Akili_shutdown2024, PEAR_bankruptcy2023}.

Notably, regulators and medical researchers are actively working on more practical standards for digital and AI mental health tools (\textit{e.g.},~\cite{apa_digital_badge}). Nevertheless, creating non-clinical LLM tools that meet clinical standards remains a difficult pathway.

A second approach to responsible design is integrating detection-and-warning features. For example, ChatGPT monitors for signs of suicidal ideation and, if detected, directs users to suicide crisis hotlines in their country. It also detects long conversations and sends ``\textit{gentle reminders to encourage breaks}''~\cite{openai_reminder_long_convo}.

However, public discourse often criticizes such detection and warning features as insufficient. 
With over a million people talking to ChatGPT about suicide \textit{each week}~\cite{million-suicidal-chatGPT-users-weekly}, providing only a phone number seems inadequate. 
Gentle reminders to take a break also appear insufficient to prevent these tools from displacing social connections or clinical care, particularly given their creators' financial interest in retaining users~\cite{YangGangCHI24_MAMH}.

\myparagraph{Section summary.~}
Prior research has identified a wealth of potential benefits and harms that LLM mental well-being support tools can create. But how designers of these tools should prioritize and act on them remains a difficult question.
Industry experience has shown that holding these tools to clinical standards is very difficult, while simple reminders to users about potential harms are inadequate. What, then, is a more practical standard for the responsible design of these tools? This paper investigates this question.



\begin{table*}[ht]
    \centering
    \renewcommand{\arraystretch}{1.3} 
    \resizebox{1\linewidth}{!}{%
        \begin{tabular}{p{0.1\linewidth} p{0.75\linewidth} p{0.1\linewidth} p{0.05\linewidth}}
        \toprule[2pt]

      \cellcolor[HTML]{f0f0f0}{\textbf{\#}} & \multicolumn{1}{c}{\cellcolor[HTML]{f0f0f0}{\textbf{AI and Human-Centered Tech Design Experts}}} &\cellcolor[HTML]{f0f0f0}{\textbf{Prof. Exp.}} & \cellcolor[HTML]{f0f0f0}{\textbf{Stage}} \\
           \midrule
           T01 & Founder of a virtual healthcare service startup. & 30+ & 1 \\ 
           T02 & Advisor to biomedical startups, former founder. & 10 - 15 & 1 \\ 
           T03 & Human-centered health tech researcher, former health tech startup founder. & 30+ & 1 \& 3 \\ 
           T04 & Advisor to health tech startups, former founder. & 15 - 20 & 1 \\ 
           T05 & Founder of a mental health tech company. & 5 - 10 & 1 \\  
           T06 & Founder COO of a mental health sensing company. & 15 - 20 & 1 \\ 
           T07 & Senior director of a consumer tech company for virtual care. & 10 - 15 & 1 \\ 
           T08 & Director of UX research at a telehealth company; & 10 - 15 & 1 \\ 
           T09 & Digital health researcher. & 5 - 10 & 1 \\ 
           T10 & AI and health sensing researcher. & 5 - 10 & 1 \\ 
           T11 & Digital health professor, working on mental health tech. & 5 - 10 & 1 \& 3 \\ 
           \midrule
        \cellcolor[HTML]{f0f0f0}{}   & \multicolumn{1}{c}{\cellcolor[HTML]{f0f0f0}{\textbf{Ethics and Policy Experts}}} & \cellcolor[HTML]{f0f0f0}{} & \cellcolor[HTML]{f0f0f0}{} \\
           \hline
           P01 & Legal and social science scholar, specializes in the impact and ethics of digital health tech. & 5 - 10 & 1 \\ 
           P02 & Data ethics and privacy scholar.  & 10 - 15 & 1 \\ 
           P03 & Health policy professor and department chair, specializes in studying the impact of technology on population health. & 20 - 25 & 1 \\ 
           P04 & Policy professor, specializes in studying the ethics of emerging technologies. & 5 - 10 & 1 \\ 
           P05 & Health policy and economics professor at a medical school, advisor to a federal regulatory agency. & 15 - 20 & 1 \& 3 \\ 
           P06 & Law professor, specializes in legal design. & 10 - 15 & 1 \\ 
           P07 & Tech policy professor, advisor to multiple federal agencies. & 25 - 30 & 1 \\ 
    
          \midrule
     \cellcolor[HTML]{f0f0f0}{}    & \multicolumn{1}{c}{\cellcolor[HTML]{f0f0f0}{\textbf{Clinical Experts}}} &\cellcolor[HTML]{f0f0f0}{} &\cellcolor[HTML]{f0f0f0}{} \\
           \hline
           C01 & Licensed Clinical Social Worker (LCSW), specializes in youth/young adult mental health, crisis intervention, and serious and persistent mental illness (SPMI) & 15 - 20 & 3  \\
           C02 & Psychiatric-Mental Health Nurse Practitioner & 10 - 15  & 3 \\
           C03 & Licensed Clinical Social Worker (LCSW-R), specializes in psychological trauma intervention & 20 - 25 & 3  \\
           C04 & Biomedical data science professor, specializes in digital and AI interventions in psychiatry & 10 - 15 & 3  \\
           C05 & Clinical psychologist and medical professor, specializes in digital mental health interventions  & 15 - 20 & 3  \\
           C06 & Medical professor, leading digital and LLM-based mental health initiatives in hospital systems & 10 - 15 & 3  \\
        \bottomrule[2pt]
        \end{tabular}
    }
    \ifTwoColThen{\vspace{0.05cm}}
    \caption{Study participants. ``Prof. Exp.'' lists participants' professional experience in years.}
    \ifTwoColThen{\vspace{-0.6cm}}
    \label{tab:participant}
\end{table*}



\section{Method}

We wanted to investigate what it means to design non-clinical LLM tools for mental well-being responsibly.
Specifically, we wanted to find a more \textit{achievable} framing of responsibility that can guide the tools' design and evaluation actions.
Given the limited prior research, we chose to investigate this question by posing it directly to experts in related fields.
We hoped by offering their answers, this work could jump-start a broader HCI community-wide discussion and surface previously unknown disagreements or questions for debate.

\subsection{Stage 1: Expert Interviews}

\myparagraph{Recruitment.~}
We sought to recruit experts whose work focuses on studying or evaluating the responsible design of AI mental well-being support (\textit{e.g.}, senior researchers in responsible AI for digital therapeutics, experienced medical ethicists, those involved in federal AI regulation, health policymakers related to AI), because they are likely to have the most relevant insights and experience for our research questions.
To achieve this, we began with extensive searches within our professional networks and then expanded recruitment through snowball sampling.

This process resulted in $18$ expert interviewees.\footnote{We intentionally did not recruit patients or users because our focus was on how designers might address benefits and risks of LLM mental well-being support that are not yet fully observable or measurable.
Interviewing users about their subjective experiences of using existing LLM tools is not suitable for this purpose. Such interviews can be found elsewhere in recent HCI research (\textit{e.g.}, \cite{YangGangCHI24_MAMH, Glossy_easy_futures,chandra2025lived}). \S\ref{RW-risk-list} summarized their valuable findings.} Among them, $12$ had over ten years of relevant experience; all had at least five years~(Table~\ref{tab:participant}).\footnote{Many human-centered design and AI participants came from startups---where much non‑clinical LLM product innovation occurs~\cite{omdenaMentalHealth,bhbusinessTalkspacePlans,SlingshotAI2025,aimmediahouseCouldThis}---and were more willing to speak with us than experts at major hospital systems or insurance companies.}

\myparagraph{Interview Protocol.~}
We designed our interview protocol with two goals in mind: (1) to gather experts’ holistic insights into what it means to design LLM mental well-being support tools responsibly, seeking convergence in their views, and (2) to ground their answers in their past actions, ensuring the insights would be actionable.

The protocol began by asking participants to recall how they prioritized and weighed the risks and benefits of such tools in their work. 
As they described their past experiences, we probed further to understand whether their actions aligned with their abstract ideals of responsible design.
For any benefit or risk they did not mention spontaneously, we presented our list of relevant harms, risks, and benefits (§\ref{RW-risk-list}) to probe responses.
All interviews lasted approximately 60 minutes; all were IRB‑approved, recorded with consent, transcribed, and de-identified.

\myparagraph{Interview Data Analysis.~}
Following the procedure of inductive thematic analysis~\cite{Braun2006Using}, two authors coded all transcripts to identify how interviewees framed responsible design and described its practical implementation. A third author reviewed the codes. Finally, three authors collaboratively sorted them into themes on a digital whiteboard, in search of novel insights.



\begin{figure*}[ht]
    \includegraphics[width=0.9\textwidth]{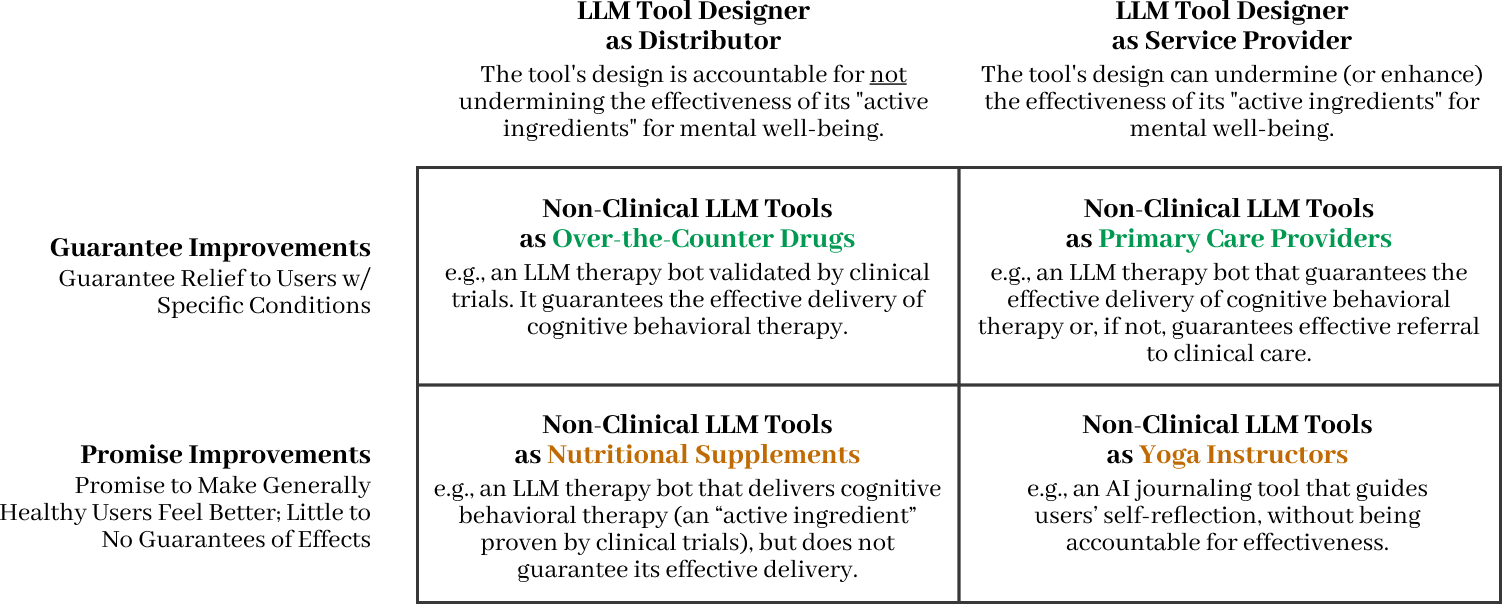}
    \caption{Our findings distinguish four types of non-clinical LLM tools based on their guaranteed benefits. This distinction has important implications for understanding what it means to design a non-clinical LLM tool responsibly.}
    \label{fig:2by2} 
\end{figure*}



\myparagraph{Emerging Findings from Stage 1.~}
Initial interviews revealed little convergence in how experts conceptualized the responsible design of LLM mental well-being support. Health policy experts considered the harms posed by these tools irrelevant to their expertise. They expressed confusion about why we sought to interview them.

\begin{quote}
    ``\textit{It is just hard to believe that the public [was] going to perceive this type of harm in the same way as the harm of, say, taking a piece of harmful medication.}'' (P03)
\end{quote}

Technology policy and responsible AI experts expressed concerns about LLM tools' impact, but did not suggest actionable solutions.
\begin{quote}
    ``\textit{There’s a culture of innovation in the US, and that serves the US well in a lot of respects. But [that meant] the evidentiary standards for regulatory action are quite high.
    [...] But at the same time, I’ve been surprised how easy it is to bring nicotine products to market.}'' (P07, tech policy professor, former federal tech policy advisor)
\end{quote}

\begin{quote}
    ``\textit{[Responsible design of LLM tools is] sort of a new territory. [...] The old frameworks just didn't apply as well. Privacy and data concerns became a big area of priority because, I think, it's an area where it still fits closely enough into existing frameworks.}'' (P01, digital therapeutics ethicist)
\end{quote}

One seemingly simple insight, however, resonated across all participants: \textit{Some LLM tools are analogous to nutritional supplements for mental well-being, while others are more like over-the-counter drugs.}
These analogies provide two distinct and accessible ways of framing the benefits and harms of non-clinical LLM tools: the drug analogy implies holding these tools to the clinical standards of an AI medical device, while the nutritional supplements analogy offers an alternative.

However, although thought-provoking, these two analogies alone do not fully answer our research questions. 
For example, what does it mean to design nutritional-supplement-like LLM tools responsibly? While nutritional supplements and drugs are fixed products, LLM apps are interactive technologies. Does this difference lead to different responsibilities?
---These questions motivated stage 2 of this research. Specifically, the first question led us to analyze regulations for nutritional supplements and over-the-counter medications to understand their distinctive responsibilities and how these are operationalized.
The second question prompted us to expand this analysis to include non-clinical mental healthcare \textit{services}, such as primary care. 

\subsection{Stage 2: Policy Analysis}

Stage 1 has revealed two useful analogies for framing the responsible design of non-clinical LLM tools: LLM tools as nutritional supplements and as over-the-counter medication. 
In stage 2, we first examined how existing U.S. regulations define and enforce responsible design for nutritional supplements, over-the-counter drugs, and comparable non-clinical healthcare services (\textit{e.g.}, primary care). We then analyzed how these approaches may or may not apply to LLM tools.

\myparagraph{Data Collection.~}
We wanted to identify existing U.S. regulations that define and enforce the responsible design of nutritional supplements, over-the-counter drugs, and comparable non-clinical healthcare services (\textit{e.g.}, primary care, school mental health counseling).\footnote{While we acknowledge the limited geographic scope, restricting our analysis to U.S. regulations allowed deep analysis of a jurisdiction where non-clinical LLM tools are regulated as consumer technology rather than medical devices.}
We began by collecting policy documents from major regulatory bodies (\textit{e.g.}, FDA, FTC, HRSA, CMS\footnote{HRSA: Health Resources and Services Administration. CMS: Centers for Medicare and Medicaid Services.}) as well as relevant legislation (\textit{e.g.}, HIPAA, DSHEA\footnote{HIPAA: Health Insurance Portability and Accountability Act of 1996. DSHEA: Dietary Supplement Health and Education Act of 1994.}), case law, and clinical guidelines.
We then expanded our search as our analysis deepened.


For example, we noticed that distinctions between products and services, as well as between those that do or do not promise health outcomes, are important in how regulators assess responsible design~(Figure~\ref{fig:2by2}). Nutritional supplements are \textit{products} that do \textit{not} guarantee health improvements; drugs are \textit{products} that do; and primary care is a \textit{service} that does. Each is held accountable in very different ways. This led us to ask: What is a service that does not promise health outcomes? How is it held accountable? These questions led us to collect and analyze regulations around yoga instruction.

\myparagraph{Data Analysis.~}
We eventually collected and read through over $100$ policy documents (Appendix~\ref{appendix_policy_reviewed}), from which we extracted more than 89 pages of notes on responsible design definitions and enforcement rules potentially transferable to LLM tools.
Three researchers, each with over five years of experience in responsible AI, law, and policy, collaboratively synthesized these definitions and rules into a uniform, actionable, and achievable framework for LLM tool responsible design. Over two months, we met multiple times per week to critique emerging frameworks and propose improvements.

\subsection{Stage 3: Deliberating Findings with Experts} \label{stage-3}

Finally, we presented our tentative findings to the experts (as defined in stage 1) and worked with them to refine and finalize these results. Our goal for this stage was to ensure that the experts could support, add nuance to, or critique any findings presented, whether from stage 1 or 2. The findings presented in the next chapter reflect their consensus, unless otherwise noted.

\myparagraph{Recruitment.~}
To ensure diverse perspectives, we recruited new experts and re-invited experts from stage 1.
When recruiting new participants, we oversampled mental health clinicians and researchers, as their expertise on the applicability of the analogies from stages 1 and 2 to mental health was particularly valuable. Among the stage 1 participants who responded, we selected a subset to ensure a balanced disciplinary background.

This process led to nine hour-long expert interviews (three returning and six new participants), as well as two 30-minute informal conversations with two additional returning participants (Table~\ref{tab:participant}).

\myparagraph{Interview Protocol.~}
To ensure all participants felt comfortable supporting, adding nuance to, or critiquing any emergent findings, we explicitly invited all three types of input during the interviews and ensured all interview questions were open-ended rather than confirmatory.
Additionally, as in stage 1, we encouraged participants to relate their opinions to their past experiences.

\myparagraph{Interview Data Analysis.~}We followed the same data analysis procedures as in stage 1.

\section{Findings}

We organize our findings on ``\textit{what it means to design non-clinical LLM tools for mental well-being responsibly}'' around three themes: (1) specific, guaranteed benefits for intended users, (2) guaranteed effective delivery of proven active ingredients, and (3) commensurate risks and benefits. We use four running analogies---non-clinical LLM tools as nutritional supplements, over-the-counter drugs, yoga instructors, and primary care providers—to illustrate the nuances of each of these responsible design criteria.



\begin{table*}[hb]
    \centering
    \renewcommand{\arraystretch}{1.3} 
    \resizebox{1\linewidth}{!}{
    \begin{tabular}[c]{p{0.07\linewidth} p{0.25\linewidth} | p{0.14\linewidth} p{0.14\linewidth} p{0.14\linewidth} p{0.12\linewidth}}
    \toprule[2pt]
    \cellcolor{mygray}   &\cellcolor{mygray}  & \multicolumn{4}{c}{\cellcolor{mygray}\textbf{Types of LLM Mental Well-Being Support}}
       \\[1.5ex] 
       \midrule
       \cellcolor{mygray}     &\cellcolor{mygray}
            &\cellcolor{mygray} Analogous to\newline \textbf{Yoga\newline Instructors}
            &\cellcolor{mygray} Analogous to\newline \textbf{Nutritional\newline Supplements}
            &\cellcolor{mygray} Analogous to\newline \textbf{Primary\newline Care}
            & \cellcolor{mygray}Analogous to\newline \textbf{OTC\newline Medication}
           \\[2ex] 
    \midrule[2pt]
        \multicolumn{2}{l|}{\textbf{Intended User Population}}
            & \multicolumn{2}{l}{Generally Healthy Individuals}
            & \multicolumn{2}{l}{Individuals w/ Specific Conditions}
            \\[2ex] 
    \midrule
        \multicolumn{2}{l|}{
                \shortstack[l]{
                    \textbf{Guaranteed Benefits}\\
                    \textbf{for Mental Well-Being}
                }
            }
            & \multicolumn{2}{l}{
                    \shortstack[l]{Minimal Guarantee\\(Promise of General Improvement)}
                }
                
            & \multicolumn{2}{l}{\shortstack[l]{Guaranteed Relief of\\ Specific, Non-Acute Conditions}}
            \\\cline{5-6}
        & & & & 
        Guaranteed\newline Effective\newline Referral
        & No Referral to \newline Specialty Care
       \\
    \midrule
       \multirow{7}{*}{\shortstack[l]{\textbf{Primary}\\\textbf{Risks}}}
       & Safety risks &  $\triangle$ Relevant &  $\triangle$ Relevant & $\blacktriangle\blacktriangle$ Top Priority & $\blacktriangle\blacktriangle$ Top Priority \\
       & Ineffectiveness & - Less Relevant & - Less Relevant & $\blacktriangle\blacktriangle$ Top Priority & $\blacktriangle\blacktriangle$ Top Priority \\
       & Displacing clinical care
        & $\blacktriangle\blacktriangle$ Top Priority
        & $\blacktriangle\blacktriangle$ Top Priority
        &  - Less Relevant
        &  $\triangle$ Relevant
        \\
       & Displacing necessary self-care
        & $\blacktriangle\blacktriangle$ Top Priority
        & $\blacktriangle\blacktriangle$ Top Priority
        &  $\triangle$ Relevant
        &  $\triangle$ Relevant
        \\
       & Risks of Goop-ifying Mental Health
        & $\blacktriangle\blacktriangle$ Top Priority
        & $\blacktriangle\blacktriangle$ Top Priority
        & - Less Relevant
        & - Less Relevant \\
       & Unequal access
        & $\triangle$ Relevant
        & $\triangle$ Relevant
        & $\blacktriangle\blacktriangle$ Top Priority
        & $\blacktriangle\blacktriangle$ Top Priority
        \\
       & Distressing comments
        & $\blacktriangle$ Important
        & $\blacktriangle$ Important
        &  - Less Relevant
        &  - Less Relevant
        \\
       \midrule
        \multicolumn{2}{l|}{\shortstack[l]{\textbf{Current Line Between Tool Creators'}\\ \textbf{and Adult Users' Responsibility}}}
            & \multicolumn{2}{p{0.28\linewidth}}{\shortstack[l]{Tool creators: Responsible for providing\\honest \& complete information.\\Users: Responsible for their choices.}}
            & \multicolumn{1}{l}{\shortstack[l]{Tool creators:\\ Responsible for \\care \& referral.}}
            & \multicolumn{1}{l}{\shortstack[l]{Same as\\supplement-like\\LLM tools.}}
            \\
    \bottomrule[2pt]
    \end{tabular}
    } 
    \vspace{0.3cm}
    \caption{The experts we interviewed shared that each type of LLM tool (as defined by its guaranteed benefits and intended users) carries distinct primary risks. This table shows the risks they considered important. They emphasized that responsible design requires proactively addressing these risks and accountability for outcomes.
    }
    \label{tab:hero}
\end{table*}



\subsection{Guaranteed Benefits for Intended Users}
For the experts we interviewed, responsible design depends on which specific benefits a tool guarantees to deliver, and to whom. Broad claims that an LLM tool ``\textit{benefits mental well-being}'' or ``\textit{alleviates physician burden}'' were insufficient. Instead, participants asked: Is the tool accountable for providing the relief a person with specific symptoms or diagnoses seeks (like an over-the-counter drug), or does it simply promise to improve the well-being of generally healthy people, without guaranteeing any immediate effects (like a nutritional supplement)? This distinction shaped how participants interpreted the responsibilities LLM tools carry and what it means to design them responsibly.

\begin{quote}
    A therapist and crisis counselor's (C02) characterization of nutritional-supplement-like LLM tools: ``\textit{(Chatbots that offer) interventions that are very low risk, and therefore generally considered positive. I will provide an example of sleep, hygiene, mindfulness, exercise, things like this. They are generally helpful for almost anybody.}'' 
\end{quote}

\begin{quote}
    FDA's definition of nutritional supplements~\cite{USC_Definitions}: \emph{Products [...] intended to supplement the diet that bear or contain one or more dietary ingredients [...] such as a vitamin; a mineral; [...] They are \emph{not} intended to diagnose, treat, cure, or prevent any disease.}
\end{quote}

\begin{quote}
    FDA's definition of pharmaceutical drugs~\cite{FDA_drug_classification}: \textit{Articles intended for use in the diagnosis of disease or other conditions, or in the cure, mitigation, treatment, or prevention of disease}. 
\end{quote}

\myparagraph{Implications for Responsible Design and Evaluation.~}
Although both types of LLM tools are \textit{non-clinical} (\textit{i.e.}, intended for use without a clinician’s prescription or supervision), participants described supplement-like and medication-like LLM tools as involving very different primary risks and, consequently, carrying very different responsibilities (Table~\ref{tab:hero}).

Participants considered the primary risks and responsibilities of over-the-counter-medication-like LLM tools to be safety and effectiveness: These tools must safely deliver guaranteed relief for specific symptoms or diagnoses to the intended users. They must ``\textit{provide what would be a clinical level of oversight and treatment.}''
\begin{quote}
    ``\textit{There are standards for all of medicine that apply well here (LLM tools claiming relief of specific symptoms or benefits), which is just evaluating risks and benefits and deployed interventions.}''~(C04)
\end{quote}

Notably, clinician and responsible AI interviewees emphasized that accessibility and equitable distribution of medication-like LLM tools are integral to their safety and effectiveness responsibilities.
\begin{quote}
``\textit{How does that (tool) influence the quality of health care for patients and consumers who do not have access to this technology? Does it further exacerbate the health equity gap, or is that actually mitigating the gap?}''~(P01)
\end{quote}
If the answer to these questions was no, several participants stated that medication-like LLM tools could not be considered safe and effective, because they would not be safe and effective for all demographics.
This perspective echoes U.S. regulations that require equitable distribution of medications, especially during public health emergencies~\cite{morgan2023advancing,american2022ensuring,cms_health-equity}.

As long as safety, effectiveness, and health equity were guaranteed, participants did not express other concerns that a medication-like LLM tool might, for example, render an overly rosy public perception of mental illnesses. Some described it as a positive if such tools help reduce stigma against these illnesses.

In sharp contrast, participants considered the primary risks of nutritional-supplement-like LLM tools to be entirely different. For them, the clinical effectiveness or ineffectiveness of these tools was less concerning than the other risks identified below.

\begin{itemize}[leftmargin=*]
    \item Individuals who are not the intended users knowingly or unknowingly use nutritional-supplement-like tools as a substitute for clinical care. ``\textit{It is less of a concern when users talk to a therapy chatbot, whether that is effective or not. It only becomes a concern when the person is talking to the chatbot when they really should talk to a psychiatrist.}'' (P05) Clinician interviewees highlighted the health risks involved in such situations; for example, therapy bots might worsen obsessive thinking in users who have undiagnosed obsessive-compulsive disorder (C02).

    \item Users use nutritional-supplement-like tools as a substitute for necessary self-care and social support. Multiple clinician interviewees expressed concerns that LLM tools might replace rather than enhance users’ social support. One clinician interviewee (C04) shared that one of their first steps when starting counseling with a new patient is to talk to the patient’s family, friends, and (if applicable) school teachers. This process includes obtaining HIPAA- and FERPA-related disclosure paperwork as needed. This allows the clinician to assess whether the patient has a proper support network outside of counseling and, if necessary, to enlist their help in the future. Because LLM mental well-being tools do not offer such care coordination, C04 and other participants considered them to introduce additional risks.

    \item Users overuse nutritional-supplement-like tools, leading to harm. ``\textit{Like supplements, if you use them (supplement-like LLM tools) too much [...] they'll damage something. For supplements, it's often the dose that will actually do harm; if you use them too much, they will damage your liver.}''\footnote{High doses of certain dietary supplements are a known cause of hepatotoxicity (liver injury)~\cite{navarro2017liver}.}~(T03)
\end{itemize}

As a result, participants considered actively preventing these risks---risks unrelated to clinical effectiveness---to be a primary responsibility of designing nutritional-supplement-like LLM tools. 
Industry participants (\textit{e.g.}, T03, T04, T06, T11) particularly stressed the need for evaluation measures that explicitly monitor for such risks and harms, noting that these issues often conflict with the tools' financial interests. As with nutritional supplements, digital mental well-being tools may profit from misleading users into believing they offer clinical benefits. Rejecting users who are not the intended audience, or who may overuse these tools, can run counter to the companies' financial incentives.

\begin{quote}
    ``\textit{Patients are rarely the customer (of digital therapeutic software). You care about the patient experience to the extent that it will benefit those real customers in the end...So I'm gonna stop there.}''~(T06) 
\end{quote}

\myparagraph{Implications for Distinguishing LLM Tool Creators' and Users' Responsibilities.~}
Just as the FDA and FTC have detailed requirements (and even design templates) to ensure accurate labeling and truthful advertising of nutritional supplements and over-the-counter drugs~\cite{FTC_health_products_guidance}, the experts we interviewed considered the creators of supplement-like and medication-like LLM tools responsible for ``\textit{giving users information and full knowledge}'' of these tools. 
Most interviewees extended this analogy further, considering these tools’ adult users responsible for their own choices based on the information provided, including overuse and misuse. Others, however, expressed reservations (more on this disagreement later, in \S\ref{finding-commensurate}).

\subsection{Guaranteed Delivery of Active Ingredients}

All experts we interviewed, in one way or another, highlighted the notion of “\textit{active ingredients}”: the mechanisms by which an LLM tool delivers its guaranteed mental well-being benefits. Clinician and health policy interviewees called this “\textit{active ingredients},” some ethicist interviewees referred to it as the tool's “\textit{theory of change,}” and industry interviewees called it the “\textit{actual essence of the product.}” 
Participants used this notion to distinguish three types of LLM tools:

\begin{enumerate}[leftmargin=*]
    \item LLM tools that have been proven effective as a whole, such as a chatbot validated through randomized controlled trials;
    \item LLM tools that deliver proven active ingredients, but the tools themselves have not been validated by randomized controlled trials. For example, an unvalidated chatbot that provides CBT;
    \item LLM tools that do not articulate their active ingredients, such as off-the-shelf ChatGPT when users use it to de-stress.
\end{enumerate}

Given that LLM tools that fall into the first category are rare (as discussed in~\S\ref{RW-design-action-challenges}), the experts we interviewed focused on the second and third types of LLM tools.

Participants consistently regarded LLM tools that cannot articulate their active ingredients as risky, and none described them as examples of responsible design.
They argued that without understanding how LLM tools “\textit{provide fun}” (\textit{i.e.}, make users feel better without a proven mechanism), it is difficult to monitor their risks and harms, potentially leading to severe, unintentional consequences.
\begin{quote}
    ``\textit{If you're using an LLM versus something that's like a branching logic type chatbot, you don't really know exactly how the bot is going to respond to something that a person might say to it, so I think that changes the safety calculation a bit.}'' (C05)
\end{quote}
Several participants (T04, T09, T11) pointed to social media as a cautionary parallel: Social media platforms also did not understand how they were “\textit{providing fun}” to users, only to discover later that this engagement was driven by “\textit{extreme polarization and anger.}”
\begin{quote}
    ``\textit{True scam companies are few and far between.[...] Very few product decisions are intentionally made to be a scam. I think oftentimes the scammy part of it comes out of, for lack of a better term, playing stupid games and winning stupid prizes. Because you set your metrics and your goals in a way that leads you to optimize over something that is not actually globally the best outcome. [...] You start by saying, `I want to offer a product that's free to people, because I think it should be accessible to everyone. Okay, but I need to make money. How do I make money? I can make money by advertising to them, because they'll have these needs. In order to advertise to them more, I need them to engage with this more, so that they see more ads.'[...] This is the downfall of social media, right? The way to engage with them more, [...] for social media, it's extreme polarization and anger.
    I haven't seen (this problem) in products like these digital health well-being products, but it could happen.}''~(T04) 
\end{quote}

Participants also shared that, beyond having a proven active ingredient, a more responsible LLM tool guarantees the effective delivery of those ingredients.
Our policy analysis revealed a parallel between LLM tools that guarantee effective delivery of their proven active ingredients and primary care services. When primary care providers offer a treatment, they ensure that each patient receives its benefits; otherwise, they are legally responsible for finding another effective treatment or making a referral to a specialty care provider.~\footnote{Despite general difficulties in proving malpractice in mental healthcare, primary care providers have faced multiple high‑profile, multimillion‑dollar lawsuits in which care providers were found liable for discharging patients without adequate suicide risk assessment, leading to preventable deaths~\cite{Kansas2025SuicideMalpractice,frierson2019malpractice,wadebyrdlawWrongfulDeath}.}
\begin{quote}
    CMS's definition of Primary Care~\cite{CMS_primary_care}: \textit{Services that intend to improve wellness, prevent and treat common illnesses. They often maintain long-term relationships with patients and [...] may also coordinate a patient’s care with specialists.}''
\end{quote}

In contrast, LLM tools that offer clinically validated treatments (\textit{e.g.}, CBT, Dialectical Behavior Therapy (DBT), Acceptance and Commitment Therapy (ACT), etc.), but do not guarantee effective delivery of those treatments, are more similar to yoga instructors than to primary care providers. While yoga instructors can substantially undermine (or enhance) the proven health benefits of yoga through their instruction, they offer no guarantee of benefits.\footnote{In case readers are curious, the proven ``active ingredients'' of yoga for mental health include mindfulness, physical exercises, and breath control~\cite{qi2020comparing_yoga,mccurdy2024program}.}
The experts we interviewed in stage 3 agreed with both analogies. They added that LLM journaling tools that prompt users to meditate and “\textit{control their journey,}” without guaranteeing the effectiveness of meditation, are also analogous to yoga instructors. 

\myparagraph{Implications for Responsible Design and Evaluation.~}
Most participants considered both (i) articulating the active ingredients of an LLM mental well-being support tool and (ii) ensuring their effective delivery as necessary criteria for the tool's responsible design and evaluation.

\begin{quote}
    ``\textit{What is the (mental health) problem you're trying to solve, and \emph{actual essence of the thing you're trying to get at} [This is referring to what we call active ingredients in this paper] in order to achieve that? And then you can set goals, and then you can look at it. It's a slippery slope where you don't set that. You could very much end up with a product that a lot of people want to sign up for, but could cause a lot of harm.}''~(T04)
\end{quote}

Furthermore, many clinician interviewees passionately argued that ``\textit{just asking if someone has suicidal thoughts is not enough}.'' 
Instead, they strongly advocated for including proper suicidal risk assessment and a guaranteed referral to clinical care in LLM mental well-being support tools. Clinician interviewees described these features as “\textit{non-negotiable}” for patient safety; 
tech policy experts considered them as basic a safety requirement as “\textit{putting on your seat belt and having driver airbags}”; and health policy interviewees noted that failures to do so create a “\textit{total health cost problem.}”

Specifically, clinician interviewees recommended that LLM tools should detect ``\textit{extreme behavior in any direction, overly positive or negative, three to five interactions in a row even if [they are spaced apart] over 24 hours}''~(C01), and ``\textit{anything that would escalate to an inpatient stay, that a trained cognitive behavioral therapist might escalate.}''~(T11) Upon detection, they recommended that LLM tools also conduct proper suicidal risk assessments, assess risk of harming others, and carry out safety planning protocols (C02, C03).

\myparagraph{Implications for Distinguishing LLM Model Developers' and Application Designers' Responsibilities.~}
Just as primary care providers are responsible for selecting effective mental health treatments, our participants suggested that it is the responsibility of LLM tool designers to select effective “\textit{active ingredients.}” By implication, it is the responsibility of the tool designers' “\textit{suppliers}”---such as creators of the language models used by the tool or developers of mental health treatments like CBT implemented by the tool---to demonstrate the safety, effectiveness, and potential risks of these ingredients to both LLM tool designers and the public.


\subsection{Commensurate Risks and Benefits}
\label{finding-commensurate}

In the two findings described so far, our participants highlighted that an LLM tool's specific promised mental well-being benefits determine its distinct primary risks and, therefore, dictate its corresponding primary responsibilities. This section examines how participants weighed benefits against risks---and where they disagreed.

Interviewees agreed that designing an LLM tool responsibly requires ensuring its risks are commensurate with its specific, guaranteed benefits. However, on two specific issues, participants were divided on which levels of risk (for whom) are commensurate with which benefits (for whom). Their disagreements often fell along disciplinary lines.

\myparagraph{Do mental well-being benefits for many users justify severe safety risks for a few?}
Most participants with backgrounds in medicine, health policy, and health technology answered yes to this question.
They raised breakthrough drugs as an analogy. The FDA has approved drugs that can cure a severe, otherwise untreatable mental illness, but carry life-and-death risks for a small subset of users, as long as these serious risks are clearly labeled and communicated to users as potential “side effects.” Following this reasoning, these participants considered LLM tools with comparable benefits and risks, if such tools exist, to be equally appropriate for real-world deployment. 

\begin{quote}
    ``\textit{Lamotrigine (a medication for bipolar disorder) has about a one in 1,000 chance of causing a serious skin reaction that can be fatal. So I have this conversation [with patients]: a low percentage chance of a high-severity side effect may be preferable.}''~(C02, psychiatric nurse practitioner)
\end{quote}

This reasoning suggests that these participants weighed the benefits and risks of LLM tools quantitatively at the population level. As they put it, it is “\textit{a very clear-cut way of thinking about safety and effectiveness}” (P05; similar quote from C04). In fact, one participant emphasized that hesitation to employ tools that may harm a few users has hindered progress in mental health interventions. 

\begin{quote}
    ``\textit{One of the factors that prevents progress from being made [in mental health], is just stigma [...]. In other areas of medicine, like oncology, they recognize that there is a risk in research and that not everyone is going to get better, some people might get worse. You have a process for educating people about those factors, and then they make their own choice.}''~(C05)
\end{quote}

Other interviewees, primarily from ethics and human-centered design backgrounds, expressed strong reservations about this reasoning. They questioned the appropriateness of weighing LLM tools’ benefits and harms solely at the population level. 

\begin{quote}
    ``\textit{The FDA works under the premise of population-level health, right? So they will approve a drug even if it killed some small amount of people or gave them really terrible side effects. [...] That runs the risk of putting us in a world where, well, so \texttt{character.ai} convinced this kid to kill himself. That's one person, but for most people, it's fine, and is that the best we can do? Is this notion of population-level health outcomes the best we can do when creating responsible design frameworks for AI mental health applications?}''~(T03)
\end{quote}


\myparagraph{Are nutritional-supplement-like LLM tools responsible designs?}

\noindent Participants broadly agreed that nutritional supplements have benefits and risks that are commensurate with each other: they offer minimal guaranteed benefits and entail minimal safety risks. However, interestingly, only a subset of participants viewed nutritional-supplement-like LLM tools as responsible designs, provided users do not treat them as substitutes for clinical care.

\begin{quote}
    ``\textit{There is nothing wrong with [LLM tools] being ineffective unless you can prove there is an individual harm.}''~(T03)
\end{quote}

Others disagreed sharply, arguing that limited utility alone disqualifies a nutritional-supplement-like LLM tool as a responsible design. 
Furthermore, they argued that, given that such tools are regulated lightly and their evaluation is “\textit{prominently, just the person's feedback about how it's going}” (C03), it is inevitable that something will go wrong.

\begin{quote}
    ``\textit{[For] the vast majority of (mental health) interventions, there's not a lot of potential harm. Limited utility would be the bigger concern.}''~(C02, clinician)
\end{quote}

\begin{quote}
    ``\emph{The phrase `\emph{wild west}' is often thrown around, but I actually think that a lot of this space (LLM mental well-being tools) deserves that label. [...] I have seen very little in this space that preemptively stops apps or products from coming online.}''~(P07, federal tech policy advisor)
\end{quote}

In the process of arguing that nutritional supplements—and by extension, nutritional-supplement-like LLM tools—are not responsible designs, two participants (T03, T11) began to question whether “\textit{commensurate benefits and risks}” was a sufficiently high standard for responsible design. After all, nutritional supplements meet this standard, yet these participants remained unsatisfied. 
However, during the interviews, they were unable to propose a concrete higher standard or specify what that standard would entail.

%

\begin{quote}
    T11 entertained the idea of raising disclosure requirements for nutritional-supplement-like LLM tools to reflect tobacco product warnings: the disclosure should not only be honest and complete, but also effective enough to discourage use:~``\textit{We all know Oreos are not great [...] And potato chips. We all eat them with the nutrition label right in front of us. Let's say you have a nutrition label for ChatGPT as a mental health support tool. First, it would have to be like, in front of everybody's face who's about to be using it, which OpenAI is not going to want, probably. [...] We can create labels every time someone visits ChatGPT, but if they like it anyway, and they felt good about using it yesterday, why wouldn't they? It [the nutritional label information] is regulation. But then it's not changing people's behavior.}''~(T11)
\end{quote}

\section{Discussion}

Millions of people now use non-clinical LLM tools, such as ChatGPT, for mental well-being support. These tools can offer a range of benefits, but also present a range of not-yet-fully-understood risks. 
With over a million users expressing suicidal thoughts to ChatGPT each week~\cite{million-suicidal-chatGPT-users-weekly}, and an unclear but likely large number of people using AI companions as substitutes for human relationships~\cite{chatGPT_loneliness_reduce_socialization_MIT2025}, we join a growing body of researchers in expressing concerns about the current and future impact of these tools, and in calling for better-designed tools and precautionary regulatory actions~\cite{YangCHI24_HCIPolicy}.

In this context, we investigated what it means to design LLM mental well-being support tools responsibly and how to operationalize that responsibility in design and evaluation.
Our findings revealed three initial criteria for responsible design: (1) specific, guaranteed benefits for intended users, (2) guaranteed effective delivery of proven active ingredients, and (3) commensurate risks and benefits. Four analogies---non-clinical LLM tools as nutritional supplements, over-the-counter drugs, yoga instructors, and primary care providers---helped illustrate the nuances of these criteria.

Below, we discuss how these findings can inform immediate actions for more responsible design of non-clinical LLM tools (\S\ref{discussion-design-action}). We also discuss how future HCI design research (\S\ref{discussion-design-rq}) and responsible AI research (\S\ref{discussion-evaluation-rq}) might critically examine and extend our findings, advancing collective understanding of how to harness LLMs' promise for mental well-being while minimizing their risks.


\subsection{Designing LLM Tools More Responsibly}
\label{discussion-design-action}

This research suggests three immediate actions that designers can take to identify, prioritize, and address the mental well-being risks and benefits of LLM tools. These actions are established best practices in HCI design, yet they have been curiously uncommon in the development of many of today’s LLM tools. We encourage LLM tool designers in the CHI research community to take the lead in re-adopting these actions and to set an example for industry practice.

\myparagraph{(I) When setting design goals, articulate the specific mental well-being benefits the LLM tool will reliably deliver and specify the intended user population.~}
Which quadrants of Figure~\ref{fig:2by2} does an LLM tool intend to occupy? Is it responsible for providing targeted relief to individuals with specific symptoms or diagnoses (like an over-the-counter drug), or does it intend to promote general well-being in otherwise healthy individuals without guaranteeing specific results (like a nutritional supplement or yoga instructor)?
If the tool detects a suicidal user, does it ensure effective referral to a human specialist and remain accountable for the user's safety during this process (as a primary care provider would)?
Our findings suggest that articulating an LLM mental well-being support tool’s specific, guaranteed benefits and intended users is a necessary first step toward delivering those benefits and being accountable for the associated risks (Table~\ref{tab:hero}).
Broad claims---that a tool “\textit{improves mental well-being}” for individuals or “\textit{makes mental healthcare more accessible}” and “\textit{alleviates shortages of mental healthcare providers}” for society---are insufficient.

Specifying measurable design goals and identifying target users are cornerstones of the human-centered design process~\cite{designcouncilDoubleDiamond}, yet they are curiously uncommon in the design of today’s commercial LLM tools.
This is not a coincidence: much of the excitement surrounding these tools stems from their ability to engage in almost any conversation with nearly any user, without requiring a clinical mental health diagnosis. As a result, requiring such tools to categorize users by diagnosis and articulate their capabilities for each user group can seem counterproductive.
However, as highlighted by the nutritional supplement and yoga analogies, LLM systems designed to serve everyone risk being accountable to no one. In health contexts, this lack of accountability can even endanger user safety.
For these reasons, we join our participants in urging designers to clearly and specifically state the mental well-being benefits that their LLM tools claim to provide.

\myparagraph{(II) In the early stages of the design process, thoughtfully select the LLM tool's “active ingredients” for improving mental well-being.~}
Our participants called for this action because LLM tools that make users feel better without a proven mechanism are analogous to social media: they merely “\textit{provide fun}” and could lead to major unintended societal consequences.

We also see the notion of an active ingredient as a more achievable goal for designing and evaluating the effectiveness of LLM mental well-being support. So far, industry practitioners and regulators have either held these tools to medical device standards (a very high bar to meet) or treated them as average consumer technology, resulting in a \textit{``wild west''} when it comes to effectiveness. 
Our research suggests a third option: ensuring that every LLM mental well-being support tool includes a mechanism proven to effectively improve mental well-being.\footnote{Notably, some participants we interviewed considered this “must-have-an-active-ingredient” standard too lenient, and for good reasons. Yet many existing LLM tools do not even meet this basic requirement. In this context, we believe that urging adherence to this standard remains a meaningful call to action.} For example, before designers of an LLM therapy chatbot can demonstrate its safety and effectiveness through clinical trials, they should thoughtfully choose the tool’s active ingredient (\textit{e.g.}, CBT) and clearly communicate it to users and regulators, along with evidence of its effectiveness and associated risks. Simply stating that a tool encourages self-reflection or mindfulness is not enough.

\myparagraph{(III) Evaluate what matters.~}
An LLM tool's acceptable levels of safety and effectiveness, as well as the types of risks it is likely to impose, depend on its target user population and its specific contributions to their mental well-being.
If a tool claims to “\textit{alleviate shortages of mental healthcare providers}” or “\textit{make mental healthcare more accessible,}” its function is analogous to a primary care service, making safety, effectiveness, and equal access its main responsibilities.
Conversely, if a tool claims nutritional supplement–like benefits, it is arguably more important to ensure that users do not use it as a substitute for clinical care or essential self-care (\textit{e.g.}, maintaining meaningful friendships) than to critique its potential ineffectiveness.

This might seem like an obvious distinction, yet it has been largely absent in prior HCI research. Instead, LLM tools published in prior HCI work (\textit{e.g.},~\cite{sharma-etal-2023-cognitive,ExploreSelf_Song_CHI25,GPTCoach_JorkeCHI25}) have consistently focused on evaluating safety, effectiveness (\textit{e.g.}, using mental health diagnostic questionnaires), and user experience (\textit{e.g.}, using perceived usefulness measures).
We see a clear need to further expand and tailor these evaluation criteria for different types of LLM tools. We hope the specific risks our participants identified, as listed below, can serve as a useful starting point for this effort.

\begin{itemize}[leftmargin=*]
    \item[] $\blacktriangleright$ Evaluation criteria for all LLM tools intended for, or capable of being used for, mental well-being support:
    \begin{enumerate}[leftmargin=*]
        \item \textit{Suicidal ideation handling:~}How accurate and precise is the tool’s detection of suicidal ideation? Upon identifying suicidal ideation, how effective is it at making referrals to clinical care, implementing safety planning, and ensuring user safety throughout the process?
        \item \textit{Other crisis handling:~}How effective is the tool at detecting and responding to other critical clinical conditions that exceed its handling capacity?
        \item \textit{Care escalation and misuse prevention:~}How effective is the tool at directing users who are not part of the intended audience to more appropriate sources of care?
        \item \textit{Truthful advertising and expectation management:~}How effective is the tool’s communication design in enabling users to understand its guaranteed benefits, intended user populations, limitations, and risks?
    \end{enumerate}
    
    \item[] $\blacktriangleright$ Additional evaluation criteria for LLM tools intended for users with specific mental health concerns or symptoms:
    \begin{enumerate}[leftmargin=*]
        \item \textit{Safety and effectiveness:~}How effective is the tool in measurably addressing user concerns or symptoms? How consistent is the tool’s effectiveness across different user groups?
        \item \textit{Clinical care coordination:~}Does the tool coordinate with the user’s other healthcare providers in delivering care? If not, what risks might arise, and how effectively does the tool address them?
        \item \textit{Health equity:~}What consequences might the tool create for those who cannot access it? What steps does the tool take to proactively address this issue, and how effective are they?
    \end{enumerate}
    
    \item[] $\blacktriangleright$ Additional evaluation criteria for LLM tools that promise, but do not guarantee, to improve generally healthy users' mental well-being:
    \begin{enumerate}[leftmargin=*]
        \item \textit{Clinical care displacement risks:~}Does use of the tool lead to users substituting it for clinical care? How effective is the tool at preventing such substitution?
        \item \textit{Essential self-care displacement risks:~}Does use of the tool lead to reduced essential mental health self-care, such as decreased socialization or lowered prosocial intentions~\cite{chatGPT_loneliness_reduce_socialization_MIT2025}?
        \item \textit{Public education benefits and risks:}~How accurate is the information the tool provides about mental illness and mental health treatment? How effective is the tool in reducing users' stigma toward these issues?
    \end{enumerate}

\end{itemize}

\subsection{Design Research Opportunities}
\label{discussion-design-rq}

This research also identified several immediate design research opportunities to support the responsible design of LLM mental well-being support tools. We highlight three here: (1) How can designers create, and regulators enforce, more responsible supplement–like LLM tools? (2) What knowledge and tools can make it easier to create primary-care–like LLM tools? (3) How might designers create LLM tools that fluidly and responsibly offer different mental well-being benefits to different users?

\myparagraph{How might designers create, and regulators enforce, more responsible supplement–like LLM tools?}
In the previous section, we urged LLM tool designers to, at a minimum, adopt the responsible design practices that the FDA requires of nutritional supplements: to ensure the presence of proven active ingredients for well-being, and to clearly and honestly communicate to users that these tools do not promise clinical benefits and cannot replace clinical care or essential self-care. In parallel to efforts that promote this basic requirement, it is important to ask: Is this good enough? Is today’s nutritional supplement landscape the future we want for LLM mental well-being support tools?

We see an urgent need for HCI designers to create LLM interfaces and interactions that prevent these applications from replacing essential self-care and clinical mental health care, and for policymakers to enforce such standards. Let’s be honest: adding prominent, nutritional supplement–style warnings to LLM tools like ChatGPT and Replika (\textit{e.g.}, ``Not a substitute for clinical care! Not a substitute for human connections!'') will not solve this problem alone, even if the warnings deter some vulnerable users.
More ingenious interaction design innovations are needed. HCI design researchers, with expertise in designing technologies for community mental health~\cite{lyon2025harnessing}, helping people become who they aspire to be~\cite{zimmerman2009designing}, and fostering a better balance between independence and social interdependence~\cite{bennett2018interdependence}, are well positioned to meet this need.

\myparagraph{What knowledge and tools can make it easier to create primary-care–like LLM tools?~}
LLM tools can have commensurate benefits and risks in ways analogous to nutritional supplements (offering minimal guarantees of health improvements to generally healthy individuals) or primary care (providing guaranteed health improvements to people in need). Designing tools that deliver the latter is clearly more difficult, yet likely much more important for society. How can HCI and AI research help make this easier?

We see two immediate opportunities that near-future research might explore to address this question. The first is to create an active ingredient database that systematically evaluates, documents, and compares proven mechanisms for improving mental well-being. For example, how reliably can various off-the-shelf language models detect suicide or severe depression? How often do clinically validated treatments like CBT succeed in different populations? Such a database could make it significantly easier for LLM tool designers to select appropriate, proven active ingredients for their tools.

The second opportunity is to divide and conquer, developing design patterns that can effectively address different aspects of mental health primary care. The evaluation criteria we listed in \S\ref{discussion-design-action} provide an initial list of these aspects. For example, what ways of prompting GPT models are most effective in delivering CBT or DBT treatments? What timing and interactions are most appropriate for an LLM tool to engage a user’s social support network and clinical care providers? Future design research can zoom into any of these questions to make a much-needed contribution.

\myparagraph{How might designers create LLM tools that fluidly and responsibly offer different mental well-being benefits to different users?~}
Overall, this research has highlighted the usefulness of differentiating an LLM tool’s risks and responsibilities based on its target users and intended benefits.
However, it does not address the reality that the same LLM tool may, in one moment, offer mundane daily conversations and, in the next, detect that a user is experiencing troubling thoughts of self-harm. 
How designers can create LLM tools that fluidly yet responsibly shift between different target benefits and user groups remains an important and challenging question for future research.

\subsection{Responsible AI Research Opportunities}
\label{discussion-evaluation-rq}

In parallel with efforts to operationalize and extend the framing of responsible design offered in this research, there should also be work that critically examines it. We highlight two open questions already emerging from our participants’ interviews and invite future research to further critique and improve upon our findings.

\myparagraph{Should we evaluate LLM tools based on population-level benefits and risks?}
Hypothetically, if an LLM tool can measurably improve the mental well-being of a vast number of users yet poses life-or-death risks to a small subset, do we accept those risks as merely “\textit{side effects}” of the tool, as FDA regulators sometimes do with breakthrough drugs? Importantly, if the answer is no, what alternative evaluation approaches are more appropriate? The participants in this study were divided on these questions.

Underlying these disagreements is the deeper question of whether we should evaluate LLM tools based on their quantitative, population-level safety and efficacy—that is, in the same way clinical trials evaluate drugs. We encourage future research to further deliberate and debate this issue. To jump-start this discussion, we propose a few alternative approaches for consideration:

\begin{itemize}[leftmargin=*]
    
    \item \textit{Rights-based approaches} can help expand the evaluation beyond weighing which benefits can justify which risks, to also consider which human rights should remain inviolable regardless of potential benefits~\cite{Prabhakaran2022Human};

    \item \textit{Value-based approaches}~\cite{friedman2013value} can potentially help navigate the value conflicts that LLM mental well-being support tools present. On one hand, the value of self-determination and informed consent supports allowing autonomous adults to choose, misuse, or overuse LLM tools (even when these tools could cause them harm) as long as all clinical benefits and risks are fully disclosed. On the other hand, the value of promoting individual and societal health may justify restricting harmful use.

    \item \textit{Public health approaches.~}Mental health is both an individual and a public health issue, as many mental health conditions are socially contagious~\cite{knapstad2020effectiveness}. It is worth considering whether we should evaluate the benefits and harms of LLM mental well-being support tools from a public health perspective, for example, by assessing how these tools might help or undermine a society’s ability to withstand stress tests during public health emergencies.
\end{itemize}

\myparagraph{Is delivering “\textit{commensurate mental health risks and benefits}” a sufficiently high standard for responsible LLM tool design?}
Experts we interviewed agreed that, in principle, a responsible LLM tool is one whose benefits and harms are at least commensurate. Based on this principle, an LLM tool analogous to a nutritional supplement (minimal guaranteed health benefits, minimal safety risks) could be considered as responsible as a tool that provides primary care (significant guaranteed health benefits, substantial safety risks due to treating individuals with existing conditions). Do we, HCI design and research communities, endorse both types of LLM tools as equally “responsible” designs? Our interviewees expressed strong but varied opinions. This divergence leaves a critical and complex question for future research to address.

\begin{acks}
The ideas in this paper were partially developed during the 2025 Everyday AI and Mental Healthcare Thought Summit, sponsored by the Cornell Center for Data Science for Enterprise and Society. Ned Cooper is partially supported by a Weill Cornell Medicine Seed Grant for the Intercampus Collaborative Project ``\textit{Improving the Robustness of Mobile Sensing and AI Systems for Mental Health Care.}'' Qian Yang is partially supported by Schmidt Futures’ AI2050 Early Career Fellowship.
\end{acks}

\bibliographystyle{ACM-Reference-Format}




\appendix
\section{Relevant Policies}\label{appendix_policy_reviewed}

This appendix lists the policies we analyzed in stage 2. For each analogy we: (1) identified the primary enforcement agency with jurisdiction; (2) identified the umbrella legislation establishing that agency's statutory authority; and (3) searched agency-specific databases to identify relevant materials, using snowball sampling from foundational statutes and targeted keyword searches (see \S3.2). We then conducted targeted desktop research to improve coverage.

In total, this process yielded over 100 statutes, regulations, guidance documents, and cases across the four analogies. Below, we list the policies reviewed.

\subsection{General Regulations}
In the U.S., market oversight of health products and services is shared between two main agencies: the Food and Drug Administration (FDA) (under the Department of Health and Human Services (HHS)) regulates safety, efficacy, and labeling of products making clinical claims, while the Federal Trade Commission (FTC) regulates marketing claims and enforces against deceptive practices.
We therefore reviewed foundational legislation for both enforcement agencies, which is relevant across all analogies.

\paragraph{Department of Health and Human Services (HHS)}
\begin{enumerate}[leftmargin=*, itemsep=0pt, parsep=0pt]
    \item \href{https://www.cdc.gov/phlp/php/resources/health-insurance-portability-and-accountability-act-of-1996-hipaa.html}{Health Insurance Portability and Accountability Act of 1996 (HIPAA)}
    \item \href{https://www.govinfo.gov/content/pkg/PLAW-111publ5/pdf/PLAW-111publ5.pdf}{Health Information Technology for Economic and Clinical Health Act of 2009 (HITECH) (Public Law 111--5, Title XIII--Health Information Technology)}
    \item \href{https://www.congress.gov/114/bills/hr34/BILLS-114hr34enr.pdf}{21st Century Cures Act (2016)}
    \item \href{https://uscode.house.gov/view.xhtml?path=/prelim@title42/chapter6A&edition=prelim}{Public Health Service Act of 1944 (42 U.S.C. Ch. 6A)}
    \setcounter{enumi_saved}{\value{enumi}}
\end{enumerate}

\paragraph{Federal Trade Commission (FTC)}
\begin{enumerate}[leftmargin=*, itemsep=0pt, parsep=0pt]
    \setcounter{enumi}{\value{enumi_saved}}
    \item \href{https://www.ftc.gov/legal-library/browse/statutes/federal-trade-commission-act}{Federal Trade Commission Act of 1914 (15 U.S.C. \S\S~41--58)}
    \item \href{https://www.ftc.gov/legal-library/browse/rules/childrens-online-privacy-protection-rule-coppa}{Children's Online Privacy Protection Rule of 1998 (16 CFR Part 312)}
    \setcounter{enumi_saved}{\value{enumi}}
\end{enumerate}

\subsection{Pharmaceutical and Medical Devices}

We reviewed the Federal Food, Drug, and Cosmetic Act (FD\&C Act) (which grants the FDA authority to regulate drugs and devices) to understand its requirements for proving safety and efficacy, and to contrast the pharmaceutical model of regulation with the supplement model (see \ref{app:supplements}).

\paragraph{Statutes}
\begin{enumerate}[leftmargin=*, itemsep=0pt, parsep=0pt]
    \setcounter{enumi}{\value{enumi_saved}}
    \item \href{https://uscode.house.gov/view.xhtml?path=/prelim@title21/chapter9&edition=prelim}{Federal Food, Drug, and Cosmetic Act of 1938 (21 U.S.C. Ch. 9)}
    \item \href{https://uscode.house.gov/view.xhtml?req=(title:21+section:356+edition:prelim)}{21 U.S.C. \S~356 (expedited approval for serious conditions)}
    \item \href{https://uscode.house.gov/view.xhtml?req=(title:42+section:262+edition:prelim)}{42 U.S.C. \S~262 (regulation of biological products)}
    \item \href{https://www.fda.gov/industry/medical-device-user-fee-amendments-mdufa-fees/medical-device-user-fee-and-modernization-act-2002-mdufma-pl-107-250}{Medical Device User Fee and Modernization Act of 2002 (P.L. 107-250)}
    \setcounter{enumi_saved}{\value{enumi}}
\end{enumerate}

\paragraph{Regulations---Drugs (21 CFR Subchapters C \& D)}
\begin{enumerate}[leftmargin=*, itemsep=0pt, parsep=0pt]
    \setcounter{enumi}{\value{enumi_saved}}
    \item \href{https://www.ecfr.gov/current/title-21/chapter-I/subchapter-C/part-201}{21 CFR Part 201 - Labeling}
    \item \href{https://www.ecfr.gov/current/title-21/chapter-I/subchapter-C/part-201/subpart-B/section-201.57}{21 CFR \S~201.57 - Specific Requirements on Content and Format of Labeling}
    \item \href{https://www.ecfr.gov/current/title-21/chapter-I/subchapter-C/part-202}{21 CFR Part 202 - Prescription Drug Advertising}
    \item \href{https://www.ecfr.gov/current/title-21/chapter-I/subchapter-C/part-203}{21 CFR Part 203 - Prescription Drug Marketing}
    \item \href{https://www.ecfr.gov/current/title-21/chapter-I/subchapter-C/part-205}{21 CFR Part 205 - Guidelines for State Licensing of Wholesale Prescription Drug Distributors}
    \item \href{https://www.ecfr.gov/current/title-21/chapter-I/subchapter-C/part-206}{21 CFR Part 206 - Imprinting of Solid Oral Dosage Form Drug Products}
    \item \href{https://www.ecfr.gov/current/title-21/chapter-I/subchapter-C/part-207}{21 CFR Part 207 - Establishment Registration and Listing for Human Drugs}
    \item \href{https://www.ecfr.gov/current/title-21/chapter-I/subchapter-C/part-208}{21 CFR Part 208 - Medication Guides for Prescription Drug Products}
    \item \href{https://www.ecfr.gov/current/title-21/chapter-I/subchapter-C/part-209}{21 CFR Part 209 - Requirement to Distribute a Side Effects Statement}
    \item \href{https://www.ecfr.gov/current/title-21/chapter-I/subchapter-C/part-210}{21 CFR Part 210 - Current Good Manufacturing Practice}
    \item \href{https://www.ecfr.gov/current/title-21/chapter-I/subchapter-C/part-250}{21 CFR Part 250 - Special Requirements for Specific Drugs}
    \item \href{https://www.ecfr.gov/current/title-21/chapter-I/subchapter-D/part-310}{21 CFR Part 310 - New Drugs}
    \item \href{https://www.ecfr.gov/current/title-21/chapter-I/subchapter-D/part-312}{21 CFR Part 312 - Investigational New Drug Application (IND)}
    \item \href{https://www.ecfr.gov/current/title-21/chapter-I/subchapter-D/part-314}{21 CFR Part 314 - New Drug Application (NDA)}
    \setcounter{enumi_saved}{\value{enumi}}
\end{enumerate}

\paragraph{Regulations---Biologics (21 CFR Subchapter F)}
\begin{enumerate}[leftmargin=*, itemsep=0pt, parsep=0pt]
    \setcounter{enumi}{\value{enumi_saved}}
    \item \href{https://www.ecfr.gov/current/title-21/chapter-I/subchapter-F/part-600}{21 CFR Part 600 - Biological Products: General}
    \item \href{https://www.ecfr.gov/current/title-21/chapter-I/subchapter-F/part-600/subpart-A/section-600.3}{21 CFR \S~600.3 - Definitions}
    \item \href{https://www.ecfr.gov/current/title-21/chapter-I/subchapter-F/part-601}{21 CFR Part 601 - Licensing}
    \setcounter{enumi_saved}{\value{enumi}}
\end{enumerate}

\paragraph{Regulations---Medical Devices (21 CFR Subchapter H)}
\begin{enumerate}[leftmargin=*, itemsep=0pt, parsep=0pt]
    \setcounter{enumi}{\value{enumi_saved}}
    \item \href{https://www.ecfr.gov/current/title-21/chapter-I/subchapter-H/part-801}{21 CFR Part 801 - Labeling}
    \item \href{https://www.ecfr.gov/current/title-21/chapter-I/subchapter-H/part-812}{21 CFR Part 812 - Investigational Device Exemptions}
    \item \href{https://www.ecfr.gov/current/title-21/chapter-I/subchapter-H/part-820}{21 CFR Part 820 - Quality System Regulation}
    \item \href{https://www.ecfr.gov/current/title-21/chapter-I/subchapter-H/part-830}{21 CFR Part 830 - Unique Device Identification}
    \setcounter{enumi_saved}{\value{enumi}}
\end{enumerate}

\paragraph{FDA Guidance Documents}
\begin{enumerate}[leftmargin=*, itemsep=0pt, parsep=0pt]
    \setcounter{enumi}{\value{enumi_saved}}
    \item \href{https://www.fda.gov/drugs/drug-approvals-and-databases/drugsfda-glossary-terms}{Drugs@FDA Glossary of Terms (2017)}
    \item \href{https://www.fda.gov/drugs/therapeutic-biologics-applications-bla/frequently-asked-questions-about-therapeutic-biological-products}{Frequently Asked Questions About Therapeutic Biological Products (2024)}
    \item \href{https://www.fda.gov/about-fda/histories-product-regulation/promoting-safe-effective-drugs-100-years}{Promoting Safe \& Effective Drugs for 100 Years (2006)}
    \item \href{https://www.fda.gov/media/183768/download}{How the FDA Regulates and Approves Drugs (n.d., accessed 2025)}
    \item \href{https://www.fda.gov/files/drugs/published/Providing-Clinical-Evidence-of-Effectiveness-for-Human-Drug-and-Biological-Products.pdf}{Providing Clinical Evidence of Effectiveness for Human Drug and Biological Products (1998)}
    \item \href{https://www.fda.gov/patients/fast-track-breakthrough-therapy-accelerated-approval-priority-review/fast-track}{Fast Track (2024)}
    \item \href{https://www.fda.gov/patients/fast-track-breakthrough-therapy-accelerated-approval-priority-review/breakthrough-therapy}{Breakthrough Therapy (2018)}
    \item \href{https://www.fda.gov/patients/fast-track-breakthrough-therapy-accelerated-approval-priority-review/priority-review}{Priority Review (2018)}
    \item \href{https://www.fda.gov/patients/fast-track-breakthrough-therapy-accelerated-approval-priority-review/accelerated-approval}{Accelerated Approval (2023)}
    \item \href{https://www.fda.gov/drugs/types-applications/new-drug-application-nda}{New Drug Application (NDA) (2022)}
    \item \href{https://www.fda.gov/drugs/types-applications/investigational-new-drug-ind-application}{Investigational New Drug (IND) Application (2025)}
    \item \href{https://www.fda.gov/safety/industry-guidance-recalls/recalls-background-and-definitions}{Recalls Background and Definitions (2014)}
    \item \href{https://www.fda.gov/drugs/laws-acts-and-rules/fdas-labeling-resources-human-prescription-drugs}{FDA's Labeling Resources for Human Prescription Drugs (2025)}
    \item \href{https://www.fda.gov/drugs/drug-approvals-and-databases/national-drug-code-directory}{National Drug Code Directory (2024)}
    \item \href{https://www.fda.gov/drugs/understanding-over-counter-medicines/over-counter-drug-facts-label}{The Over-the-Counter Drug Facts Label (2024)}
    \item \href{https://www.fda.gov/medical-devices/classify-your-medical-device/how-determine-if-your-product-medical-device}{How to Determine if Your Product is a Medical Device (2022)}
    \item \href{https://www.fda.gov/medical-devices/digital-health-center-excellence/digital-health-policy-navigator}{Digital Health Policy Navigator (2022)}
    \item \href{https://www.fda.gov/medical-devices/premarket-submissions-selecting-and-preparing-correct-submission/premarket-approval-pma}{Premarket Approval (PMA) (2019)}
    \item \href{https://www.fda.gov/medical-devices/premarket-submissions-selecting-and-preparing-correct-submission/premarket-notification-510k}{Premarket Notification 510(k) (2024)}
    \item \href{https://www.fda.gov/medical-devices/premarket-submissions-selecting-and-preparing-correct-submission/de-novo-classification-request}{De Novo Classification Request (2025)}
    \item \href{https://www.fda.gov/medical-devices/premarket-submissions-selecting-and-preparing-correct-submission/humanitarian-device-exemption}{Humanitarian Device Exemption (2025)}
    \item \href{https://www.fda.gov/media/82395/download}{The 510(k) Program: Evaluating Substantial Equivalence in Premarket Notifications (2014)}
    \item \href{https://www.fda.gov/medical-devices/classify-your-medical-device/device-classification-panels}{Device Classification Panels (2025)}
    \item \href{https://www.fda.gov/regulatory-information/search-fda-guidance-documents/fda-and-industry-procedures-section-513g-requests-information-under-federal-food-drug-and-cosmetic}{FDA and Industry Procedures for Section 513(g) Requests for Information (2024)}
    \item \href{https://www.fda.gov/regulatory-information/search-fda-guidance-documents/general-wellness-policy-low-risk-devices}{General Wellness: Policy for Low Risk Devices (2025)}
    \item \href{https://www.fda.gov/combination-products}{Combination Products (n.d., accessed 2025}
    \item \href{https://www.fda.gov/medical-devices/digital-health-center-excellence/device-software-functions-including-mobile-medical-applications}{Device Software Functions Including Mobile Medical Applications (2022)}
    \item \href{https://www.fda.gov/medical-devices/digital-health-center-excellence/software-medical-device-samd}{Software as a Medical Device (SaMD) (2018)}
    \item \href{https://www.fda.gov/medical-devices/software-medical-device-samd/artificial-intelligence-software-medical-device}{Artificial Intelligence in Software as a Medical Device (2025)}
    \item \href{https://www.fda.gov/regulatory-information/search-fda-guidance-documents/marketing-submission-recommendations-predetermined-change-control-plan-artificial-intelligence}{Marketing Submission Recommendations for a Predetermined Change Control Plan for AI-Enabled Device Software Functions (2025)}
    \item \href{https://www.fda.gov/medical-devices/software-medical-device-samd/transparency-machine-learning-enabled-medical-devices-guiding-principles}{Transparency for Machine Learning-Enabled Medical Devices: Guiding Principles (2024)}
    \item \href{https://www.fda.gov/regulatory-information/search-fda-guidance-documents/artificial-intelligence-enabled-device-software-functions-lifecycle-management-and-marketing}{AI-Enabled Device Software Functions: Lifecycle Management and Marketing Submission Recommendations (2025)}
    \item \href{https://www.fda.gov/medical-devices/software-medical-device-samd/artificial-intelligence-enabled-medical-devices}{Artificial Intelligence-Enabled Medical Devices (2025)}
    \item \href{https://www.fda.gov/medical-devices/overview-device-regulation/device-labeling}{Device Labeling - Regulatory Requirements for Medical Devices (2020)}
    \item \href{https://www.fda.gov/regulatory-information/search-fda-guidance-documents/device-labeling-guidance-g91-1-blue-book-memo}{Device Labeling Guidance \#G91-1 (Blue Book Memo) (2018)}
    \item \href{https://www.fda.gov/regulatory-information/search-fda-guidance-documents/guidance-medical-device-patient-labeling}{Guidance on Medical Device Patient Labeling (2001)}
    \setcounter{enumi_saved}{\value{enumi}}
\end{enumerate}

\paragraph{Case Law}
\begin{enumerate}[leftmargin=*, itemsep=0pt, parsep=0pt]
    \setcounter{enumi}{\value{enumi_saved}}
    \item \href{https://supreme.justia.com/cases/federal/us/555/555/}{\textit{Wyeth v. Levine}}, 555 U.S. 555 (2009)
    \item \href{https://supreme.justia.com/cases/federal/us/564/604/}{\textit{PLIVA, Inc. v. Mensing}}, 564 U.S. 604 (2011)
    \setcounter{enumi_saved}{\value{enumi}}
\end{enumerate}

\subsection{Primary Care Services}

Primary care is regulated as a service rather than a product, which means oversight is fragmented across multiple regulatory bodies rather than consolidated under a single agency. To reflect this patchwork of regulation, we reviewed relevant Federal statutes, State licensing laws and regulations (using New York State as a case study), clinical practice guidelines and policy documents, and case law on medical malpractice.

\paragraph{Federal Statutes}
\begin{enumerate}[leftmargin=*, itemsep=0pt, parsep=0pt]
    \setcounter{enumi}{\value{enumi_saved}}
    \item \href{https://uscode.house.gov/view.xhtml?req=(title:42+section:1395nn+edition:prelim)}{42 U.S.C. \S~1395nn (Stark Law / Physician Self-Referral)}
    \item \href{https://uscode.house.gov/view.xhtml?req=(title:42+section:1320a-7b+edition:prelim)}{42 U.S.C. \S~1320a-7b(b) (Anti-Kickback Statute)}
    \item \href{https://uscode.house.gov/view.xhtml?req=(title:42+section:1395u+edition:prelim)}{42 U.S.C. \S~1395u (Medicare Part B administration)}
    \setcounter{enumi_saved}{\value{enumi}}
\end{enumerate}

\paragraph{New York State Law and Regulations}
\begin{enumerate}[leftmargin=*, itemsep=0pt, parsep=0pt]
    \setcounter{enumi}{\value{enumi_saved}}
    \item \href{https://www.op.nysed.gov/title8/education-law/article-131}{Education Law Article 131 (Medicine, \S\S~6520--6529)}
    \item \href{https://www.health.ny.gov/professionals/office-based_surgery/law/6530.htm}{Education Law \S~6530 (professional misconduct)}
    \item \href{https://www.nysenate.gov/legislation/laws/PBH/238-A}{Public Health Law \S~238-A (prohibition on referral arrangements)}
    \item \href{https://www.health.ny.gov/facilities/cons/}{Certificate of Need requirements}
    \item \href{https://www.health.ny.gov/health_care/managed_care/pdf/subpart98-1and2.pdf}{10 NYCRR Subparts 98-1 and 98-2 (Managed Care Organizations)}
    \item \href{https://regs.health.ny.gov/content/section-10032-definitions}{10 NYCRR \S~1003.2 (definitions)}
    \item \href{https://regs.health.ny.gov/content/section-7002-definitions}{10 NYCRR \S~700.2 (definitions - medical facilities)}
    \item \href{https://www.health.ny.gov/facilities/public_health_and_health_planning_council/meetings/2019-03-28/docs/primary_care_clinic_classifications.pdf}{New York Primary Care Clinic Classifications}
    \item \href{https://www.health.ny.gov/publications/1500/}{New York State Hospital Patients' Bill of Rights}
    \setcounter{enumi_saved}{\value{enumi}}
\end{enumerate}

\paragraph{Clinical Guidelines and Policy Documents}
\begin{enumerate}[leftmargin=*, itemsep=0pt, parsep=0pt]
    \setcounter{enumi}{\value{enumi_saved}}
    \item \href{https://www.cms.gov/priorities/innovation/key-concepts/primary-care}{CMS.gov: Primary Care (2023)}
    \item \href{https://www.cms.gov/files/document/hhs-cms-glossary-health-coverage-and-medical-termspdf}{HHS/CMS Glossary of Health Coverage and Medical Terms (n.d., accessed 2025}
    \item \href{https://bhw.hrsa.gov/glossary}{Health Resources and Services Administration (HRSA) Glossary (2025)}
    \item \href{https://publications.aap.org/pediatrics/article/120/5/e1299/71095/Guidelines-for-Adolescent-Depression-in-Primary}{Guidelines for Adolescent Depression in Primary Care (GLAD-PC) (2007)}
    \item \href{https://publications.aap.org/pediatrics/article/137/Supplement_2/S136/33995/Irritability-and-Problem-Behavior-in-Autism}{Irritability and Problem Behavior in Autism Spectrum Disorder: A Practice Pathway for Pediatric Primary Care (2016)}
    \setcounter{enumi_saved}{\value{enumi}}
\end{enumerate}

\paragraph{Case Law}
\begin{enumerate}[leftmargin=*, itemsep=0pt, parsep=0pt]
    \setcounter{enumi}{\value{enumi_saved}}
    \item \href{https://law.justia.com/cases/washington/supreme-court/1974/42775-1.html}{\textit{Helling v. Carey}}, 519 P.2d 981 (Wash. 1974)
    \item \href{https://law.justia.com/cases/federal/appellate-courts/cadc/22099/22099.html}{\textit{Canterbury v. Spence}}, 464 F.2d 772 (D.C. Cir. 1972)
    \item \href{https://law.justia.com/cases/minnesota/supreme-court/2022/a20-0711.html}{\textit{Smits v. Park Nicollet Health Services}}, 979 N.W.2d 436 (Minn. 2022)
    \item \textit{Yates v. AdventHealth}, No. 22CV03467 (Kan. Dist. Ct. 2025)
    \setcounter{enumi_saved}{\value{enumi}}
\end{enumerate}

\subsection{Nutritional Supplements}\label{app:supplements}

We reviewed the Dietary Supplement Health and Education Act of 1994 (DSHEA) and relevant regulations in the Code of Federal Regulations (CFR) to understand the regulatory model for nutritional supplements and contrast it with the pharmaceutical model.

\paragraph{Statutes}
\begin{enumerate}[leftmargin=*, itemsep=0pt, parsep=0pt]
    \setcounter{enumi}{\value{enumi_saved}}
    \item \href{https://www.congress.gov/bill/103rd-congress/senate-bill/784/text}{Dietary Supplement Health and Education Act of 1994 (DSHEA)}
    \item \href{https://www.fda.gov/food/food-allergensgluten-free-guidance-documents-regulatory-information/food-allergen-labeling-and-consumer-protection-act-2004-falcpa}{Food Allergen Labeling and Consumer Protection Act of 2004 (FALCPA)}
    \item \href{https://uscode.house.gov/view.xhtml?req=(title:21+section:331+edition:prelim)}{21 U.S.C. \S~331 (prohibited acts)}
    \item \href{https://www.fda.gov/tobacco-products/rules-regulations-and-guidance/family-smoking-prevention-and-tobacco-control-act-overview}{Family Smoking Prevention and Tobacco Control Act of 2009 (P.L. 111-31)}\footnote{Reviewed because DSHEA's statutory definition explicitly excludes tobacco products (21 U.S.C. \S~321(ff)), making this act relevant to understanding regulatory boundaries.}
    \setcounter{enumi_saved}{\value{enumi}}
\end{enumerate}

\paragraph{Regulations}
\begin{enumerate}[leftmargin=*, itemsep=0pt, parsep=0pt]
    \setcounter{enumi}{\value{enumi_saved}}
    \item \href{https://www.ecfr.gov/current/title-21/chapter-I/subchapter-B/part-101/subpart-C/section-101.36}{21 CFR \S~101.36 (nutrition labeling of dietary supplements)}
    \item \href{https://www.ecfr.gov/current/title-21/chapter-I/subchapter-B/part-190/section-190.6}{21 CFR \S~190.6 (premarket notification for new dietary ingredients)}
    \item \href{https://www.ecfr.gov/current/title-21/chapter-I/subchapter-B/part-111}{21 CFR Part 111 (Good Manufacturing Practice for dietary supplements)}
    \setcounter{enumi_saved}{\value{enumi}}
\end{enumerate}

\paragraph{FDA Guidance Documents}
\begin{enumerate}[leftmargin=*, itemsep=0pt, parsep=0pt]
    \setcounter{enumi}{\value{enumi_saved}}
    \item \href{https://www.fda.gov/food/retail-food-protection/fda-food-code}{2022 FDA Food Code}
    \item \href{https://www.fda.gov/regulatory-information/search-fda-guidance-documents/guidance-industry-new-dietary-ingredient-notification-procedures-and-timeframes-dietary-supplements}{New Dietary Ingredient Notification Procedures and Timeframes (2024)}
    \item \href{https://www.fda.gov/regulatory-information/search-fda-guidance-documents/guidance-industry-substantiation-dietary-supplement-claims-made-under-section-403r-6-federal-food}{Substantiation for Dietary Supplement Claims Made Under Section 403(r)(6) (2009)}
    \item \href{https://www.fda.gov/about-fda/economic-impact-analyses-fda-regulations/clarification-when-products-made-or-derived-tobacco-are-regulated-drugs-devices-or-combination}{Clarification of When Products Made or Derived From Tobacco Are Regulated as Drugs, Devices, or Combination Products (2017)}
    \setcounter{enumi_saved}{\value{enumi}}
\end{enumerate}

\subsection{Yoga Instruction} \label{app:yoga}

At the time of submitting this article, there is no overarching federal regulation of yoga instruction in the U.S. We therefore reviewed how responsibility is managed in the absence of statutory regulation: primarily through contract law (liability waivers signed by participants) and general negligence principles.

\end{document}
\endinput